\documentstyle[12pt,fleqn]{article} 
\newcommand{\beq}{\begin{equation}} 
\newcommand{\eeq}{\end{equation}} 
\newcommand{\bea}{\begin{eqnarray}} 
\newcommand{\eea}{\end{eqnarray}} 
\newcommand{\beau}{\begin{eqnarray*}}
\newcommand{\eeau}{\end{eqnarray*}}

\begin{document}

 \begin{center}
 {\bf  Sparse random matrix configurations for two or three
interacting electrons in a random potential}\\
 {\sl Shi-Jie Xiong$^{1,2}$ and S.N. Evangelou$^{1,3}$\\
 $^1$Foundation for Research and Technology, \\
 Institute for Electronic Structure and Lasers, \\
 Heraklion, P.O. Box 1527, 71110 Heraklion, Crete, Greece \\
 $^2$Department of Physics, Nanjing University,\\
 Nanjing 210008, China\\
 $^3$Department of Physics, University of
 Ioannina,\\ Ioannina 451 10, Greece}

 \bigskip

 \end{center}
\begin{abstract} 
We investigate the random matrix configurations for two or three
interacting electrons in one-dimensional disordered systems.
In a suitable non-interacting localized electron basis we obtain
a sparse random matrix with very long tails which is different 
from a superimposed random  band matrix usually thought 
to be valid. The number of non-zero off-diagonal matrix elements
is shown to decay  very weakly from the matrix  diagonal and the  
non-zero matrix elements are distributed according to a 
Lorentzian around zero with also very weakly decaying parameters.
The corresponding random matrix for three interacting electrons 
is similar but even more sparse. 
\end{abstract}
\vspace{1cm}


\newpage

There is a great current interest in the 
localization weakening effect due to the interaction 
of two electrons in one-dimensional (1D) disordered 
systems \cite{1,2,3,4,5,6,7,8,9,10}. Shepelyansky \cite{1}
mapped this problem onto a class of random banded matrices with
strongly fluctuating  diagonal elements, being the eigenenergies of 
the non-interacting  problem, and 
independent Gaussian random off-diagonal matrix elements 
of zero average and typical strength ${\frac {U}{\xi_{1}^{3/2}}}$,
lying in a band of width  the one particle localization length 
$\xi_1$ with $U$ the strength of the interaction. 
Moreover, by the mapping to a
superimposed banded random matrix ensemble ($SBRM$)
a fraction of ${\frac {\xi_1}{2 L}}$ states with a
considerable enhancement $\xi \sim U^{2} \xi_1^{2}$ 
of the localization length along the center of mass coordinate 
is predicted
due to coherent propagation of the electronic pair. It was also
shown that the interaction has no effect for the  majority of
other states with the two particles localized in isolated spatial
positions which do not allow overlapping.
This conclusion was confirmed and extended to higher 
dimensions via  Thouless block scaling picture by Imry \cite{2}.
The subsequent numerical studies \cite{3,4,5,6,7,8,9,10} 
verified the main qualitative results concerning
the presence of Shepelyansky states, mostly by supressing 
single particle transport via efficient Green function  
or  bag model methods which examine pair propagation
\cite{8}. The deviations from the predicted behavior of the 
two-particle localization length $\xi$ found
were usually attributed to the oversimplified statistical 
assumptions concerning the band random matrix model of the
original Shepelyansky construction.

However, there is an ongoing debate whether 
coherent pair propagation actually exists for two interacting
electrons in infinite disordered systems \cite{11,12}, 
which began by a recent transfer matrix study
where  no propagation enhancement is found at $E=0$
for an infinite chain \cite{11}.  Moreover, it was pointed out
that the reduction to a $SBRM$ relies on questionable assumptions 
regarding chaoticity of the non--interacting electron 
localized states within $\xi_1$, so that the relevant matrix model
could be probably different \cite{13,14}. Although the reported
absence of propagation enhancement \cite{11} can be critisized,
since the transfer matrix method may not measure the actual
pair localization length, it is correct that the fraction
of the Shepelyansky states will eventually shrink to zero 
when the system size increases, although not affecting
their physical significance.  A different localization length
enhancement  for theses states, of the form 
$\xi \sim U^{2} \xi_1^{1+\gamma}$ with 
a $U$--dependent exponent $\gamma <1$, was also proposed 
on the basis of numerical data
by a reduction mechanism  to another appropriate 
random matrix model \cite{13}. It must be mentioned that
the Shepelyansky states are expected to exist 
as long as the interaction is not too large, since
it is  firmly established \cite{10,13} that in the strong
interaction limit $U\to \infty$ no coherent propagation 
enhancement is possible with  these states 
decoupled from the main band with $\xi \approx  \xi_1$. 

It is worthwhile to check  the validity  
of the mapping to a $SBRM$ which allowed most of the previous
results concerning the Shepelyansky states to be derived.
The fact that a mapping to  $SBRM$ neglects  phase correlations
of the one particle localized states was also emphasised 
for a few known examples in another recent study \cite {14},  
where it lead to a non--justified propagation enhancement. 
In  order to shed light on the appropriate random matrix
description for the  problem of two interacting electrons
in a random potential we examine explicitly  the structure of 
the two and three-electron Hamiltonian matrices by a 
direct numerical analysis. The corresponding Anderson-Hubbard 
Hamiltonian\cite{15,16} can be written as 
\[
H=\sum_{n=1}^N\sum_{\sigma }(c^{\dag }_{n+1,\sigma }c_{n,\sigma } 
 +\mbox{H.c.})+\sum_{n=1}^N\sum_{\sigma }
\epsilon_nc^{\dag }_{n, \sigma }
 c_{n,\sigma }
 \]
 \begin{equation}
\label{ham}
\ \ \ \ \ \ +\sum_{n=1}^N\sum_{\sigma \neq \sigma '}
U c^{\dag }_{n,\sigma '}c_{n,\sigma '}
 c^{\dag }_{n,\sigma }c_{n, \sigma },
\end{equation}
where $c^{\dag }_{n,\sigma }$ and $c_{n,\sigma }$ are the 
creation and destruction operators for the electron
at site $n$ and with spin $\sigma $,
$\epsilon_n$ is energy level at site $n$ which 
is a random variable uniformly distributed
in the range $[-W/2, W/2]$ as for the Anderson model  and 
$U$ is the strength of the interaction between the 
electrons. 

A non--interacting disordered system 
of $N$ atoms with $U=0$ in Eq. (1) has $N$ linearly independent 
one--electron localized wave functions $\psi_i$
with corresponding eigenvalues $E_i$, $i=1,2,3,...,N$. 
These exponentially localized wave functions are 
of the approximate  form
\beq
\psi_i(n) \sim {\frac {1}{\sqrt{\xi_1}}} \exp \left[-{\frac{
|n-n_i|}{\xi_1} +i \theta_{i}(n)}\right],
\eeq
where $n_i$ is the localization center of $\psi_i$ and 
$\theta_i$ a corresponding phase factor. The 
corresponding perturbational localization length for 
small $W$ is $\xi_1 \approx {\frac {96-24E^{2}}{W^{2}}}$ where $E$
is the single particle energy, although for some special $E$
the prefactor is different. In order to examine the few--body
problem in the presence of disorder we  have obtained numerically
all the localized states $\psi_i$ and arranged them so that 
if $i<j$ the coordinate $n_i$ for the localization center
of $\psi_i$ is smaller than the corresponding center $n_j$ 
of the wave function $\psi_j$. This is a natural kind of 
arrangement which quarantees the largest overlapping of the wave
functions to occur  when their indices have the smallest
difference.

Firstly, we consider two electrons with opposite spins
and use the $N^{2}$ products of the two one-electron wave
functions 
\begin{equation}
\Psi^{(2)}_m=\psi_i\psi_j \mbox{  , }  i,j=1,2,..,N,
\end{equation}
as convenient basis states for the two interacting electrons.
We do not consider the spin configuration
of the two-electron wave function as suggested in Ref. \cite{1}. 
The index $m$  is also arranged in such a way so that
states with the smallest difference of their indices have the 
strongest coupling. One can calculate the matrix elements of
the $U$--dependent Hubbard interaction term of Eq. (1) 
in the obtained  new basis set via
\beq
H_{m,m'} = U \sum_{n} \psi_i(n) \psi_j(n) 
\psi_{i'}(n) \psi_{j'}(n).
\eeq
In the original $SBRM$ construction 
the matrix element $H_{m,m'}$ was shown to
vanish  unless all four relevant wave functions were 
overlapping. Then  each $\psi_i(n)$ was
assumed completely  random within $\xi_1$ by
taking the approximate states of Eq. (2) with  a random 
phase factor $\theta_{i}(n)$, so that one immediately  
obtains the estimate ${\frac {U}{\xi_{1}^{3/2}}} $ for the 
typical magnitude of the off--diagonal matrix elements 
distributed within a band range. The $SBRM$
construction  ignores  phase correlations which are known 
to give regular or fast oscillations within $\xi_1$
for  the single particle localized wave functions.

In Figs. 1 and 2 we show the obtained distribution of the 
diagonal and the off-diagonal matrix elements for two
interacting electrons with  disorder extent  $W$ and 
interaction strength $U$. The diagonal matrix 
elements are seen to obey a Gaussian distribution as expected.
The off-diagonal matrix elements are found to be mostly zero 
but also very close to zero with a Lorentzian distribution
having  very long  tails in the latter case. In order to fit the 
obtained distribution for  the off-diagonal matrix elements $h$
in the adopted basis we use the following  function sum
\begin{equation}
f(h)=f_{G}(h)+f_{L}(h) = a_1\exp \left( -\frac{1}{a_2}h^2 \right)
 +\frac{a_3}{h^2+a_4},
\end{equation}
where the  Gaussian--like term $f_{G}$
accounts for the distribution of the off-diagonal elements
which are very close to zero and the Lorentzian--like
term  $f_{L}$  for the long tails. The variations
of the obtained four fitting parameters 
$a_1$, $a_2$, $a_3$ and $a_4$  
as a function of the distance of the matrix element locations
from the main matrix diagonal are shown to decay extremely weakly
in Figs. 3 and 4. 
It can immediately be seen that $a_1$ is much larger than 
$a_3$ which implies a very large
number  of near--zero matrix elements. The obtained 
parameters $a_2, a_4$ characterising the distribution
are found very small. 

We have also investigated the
random matrix structure for the  corresponding three 
electron problem. In order to simplify the calculations
we have restricted our consideration to three electrons where
two of them have  spin up and one spin down. 
For a chain of $N$--sites in the adopted three--electron basis
there are $N^2(N-1)/2$ linearly independent states
whose wave functions can be written as
\begin{equation}
\Psi^{(3)}_m=\psi_{i,\uparrow }\psi_{j, \uparrow }
\psi_{k, \downarrow } \mbox{  , } i \neq j.
\end{equation}
We again sort these  $3$-particle  states in such a way so
that those with the smallest difference in their indices
have the strongest possible coupling. In Fig. 5 we display the
obtained distributions for  the diagonal
and the  off-diagonal matrix elements. For the diagonal elements 
the distribution is more sharp when compared to the two-electron
case, reflecting a smearing effect in the total energy
fluctuations due to the increase in the number of particles. 
We find that the off-diagonal matrix elements obey the 
general principles of the two--electron problem but
the matrix structure is  comparatively even more sparse 
in this case. 

In summary, we have studied the random matrix configurations 
for two or three interacting electrons in a 
disordered chain. We made a basis set arrangement  
which consists of non-interacting particle
wave function products of orbitals  so that the 
off-diagonal matrix elements due to 
the interaction occur successively only for states having 
closely spaced indices. Although this is the most favourable 
basis set for obtaining a $SBRM$ a very sparse random matrix 
structure is shown to emerge  from our data instead, having
no well--defined band region with Gaussian
matrix elements. We think   finding another basis set
consistent with a band random matrix structure is not easy 
and the few-body problem can be studied via 
extremely sparse random matrix configurations. 
It must be  pointed out that the obtained sparse matrix 
is not incompatible with states having enhanced 
localization length along the center of mass coordinate, 
since they can be shown by methods which do not rely on the 
specific matrix mapping \cite {8,17}.

 {\Large\bf Acknowledgments} 

 This work was supported in part by a $\Pi$ENE$\Delta$ Research 
 Grant of the Greek Secretariat of Science and Technology,
 from EU contract CHRX-CT93-0136 and a TMR network. We also
 thank F. von Oppen for an  e-mail correspondance.

 \newpage\clearpage

 \begin{figure}[h]
 {\Large \bf Figure Captions}
 \\
  {\bf Fig. 1.}   The distribution of the 
 matrix elements for  
 the two-electron Hamiltonian with disorder $W=3$, 
 interaction strength $U=1$ and chain size $N=100$. 
 {\bf (a)}  The diagonal matrix elements. 
 {\bf (b)}  The first off-diagonal matrix elenents 
 (continuous line) and the fitting curve from Eq. (5) 
 (broken line).

 {\bf Fig. 2.}   The same as Fig. 1 but for $W=6$ and 
interaction strength $U=4$. 
 
 {\bf Fig. 3.}  The four fitting parameters $a_1, a_2, a_3, a_4$
 of Eq. (5) for the distribution of the off-diagonal matrix 
 elements corresponding to Fig. 1 as a function of the 
 distance from the main matrix diagonal, with chain size $N=100$,
 disorder $W=3$ and interaction strength $U=1$.
 
 {\bf Fig. 4.}  The same as Fig. 3 but for $W=6$ and 
 interaction strength $U=4$ corresponding to Fig. 2. 

 {\bf Fig. 5.}  The distribution of the 
 matrix elements for  
 the three-electron Hamiltonian  with disorder $W=6$, 
 interaction strength $U=4$ and chain size $N=100$. 
 {\bf (a)}  The distribution of the diagonal matrix elements. 
 {\bf (b)}  The distribution of the first off-diagonal matrix 
 elenents. 

\end{figure}
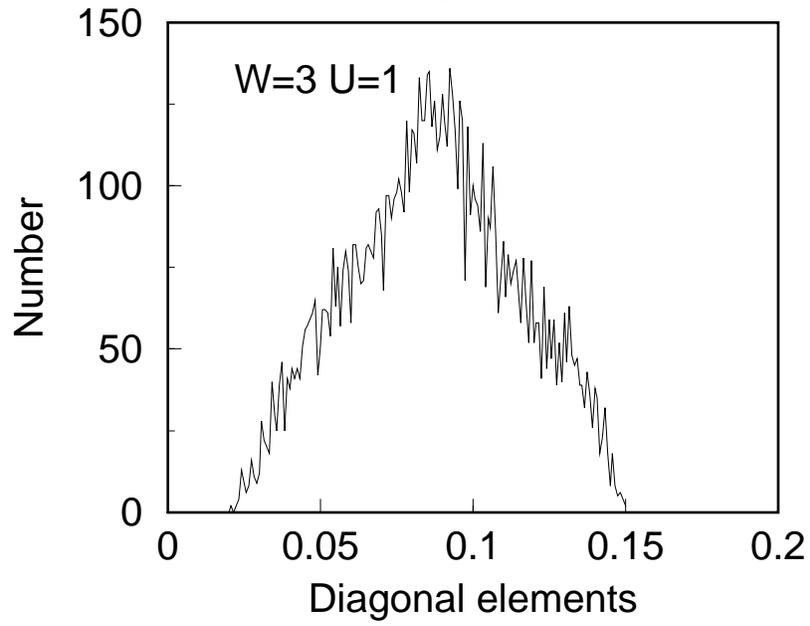
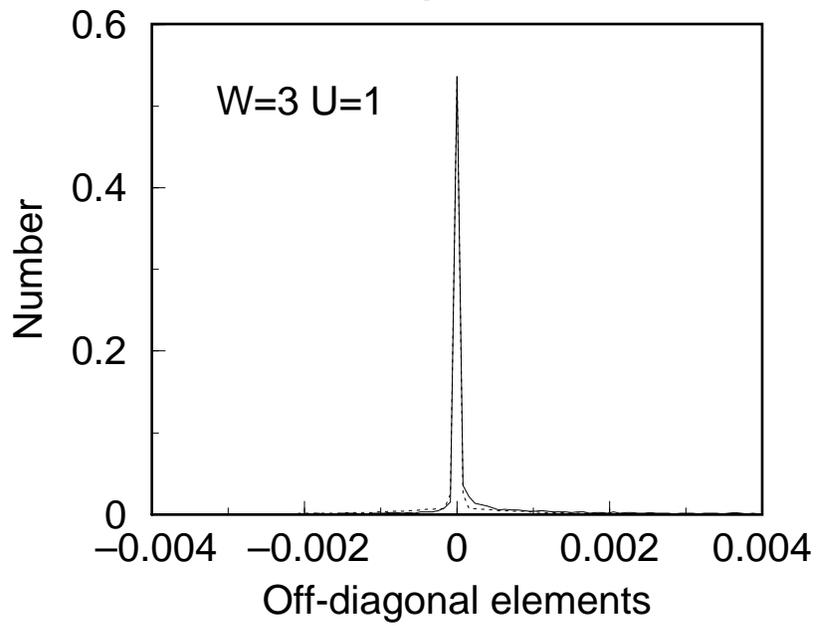
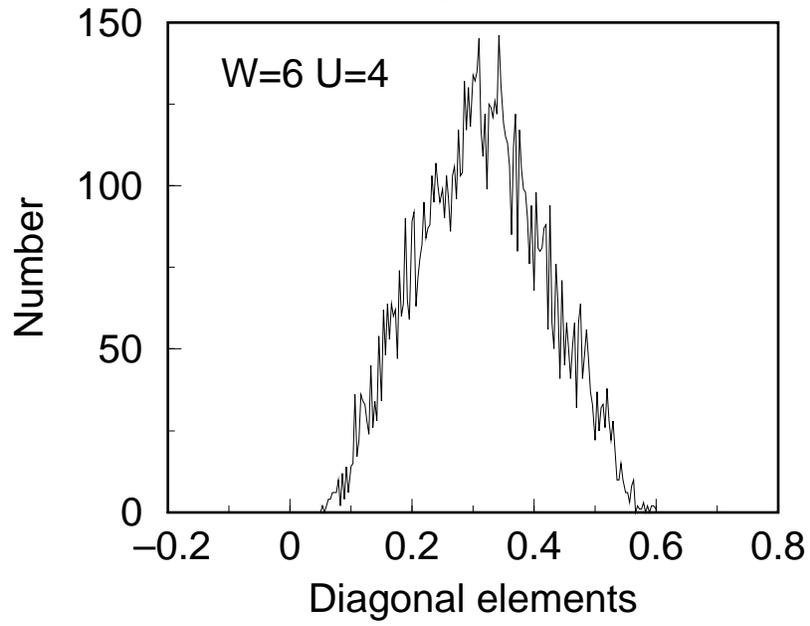
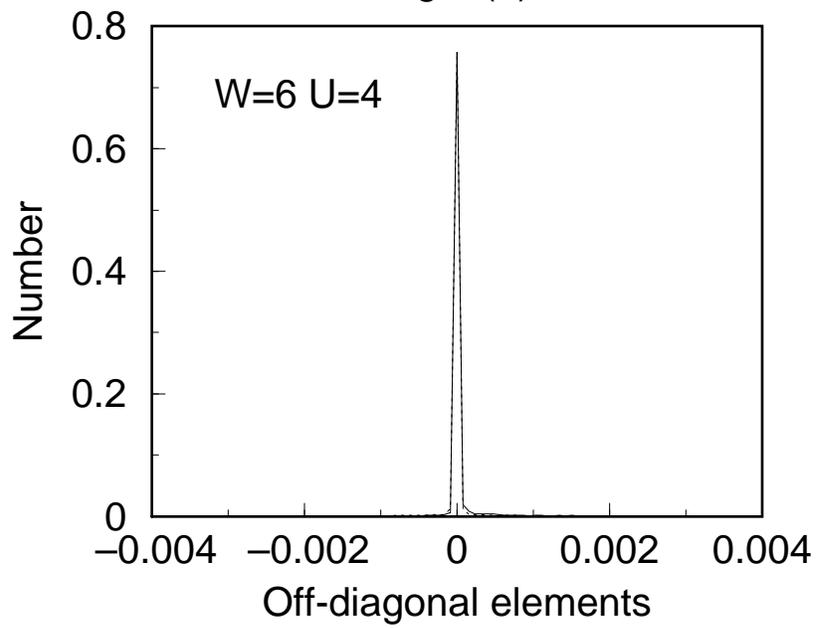
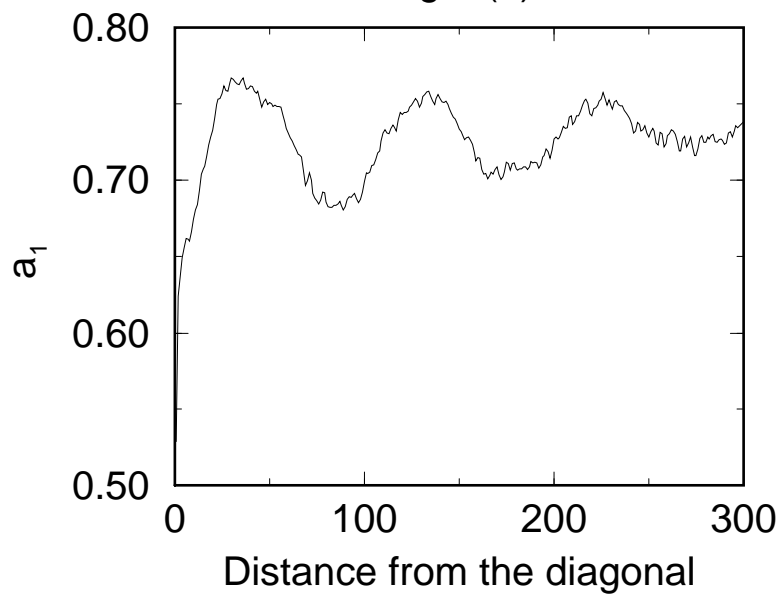
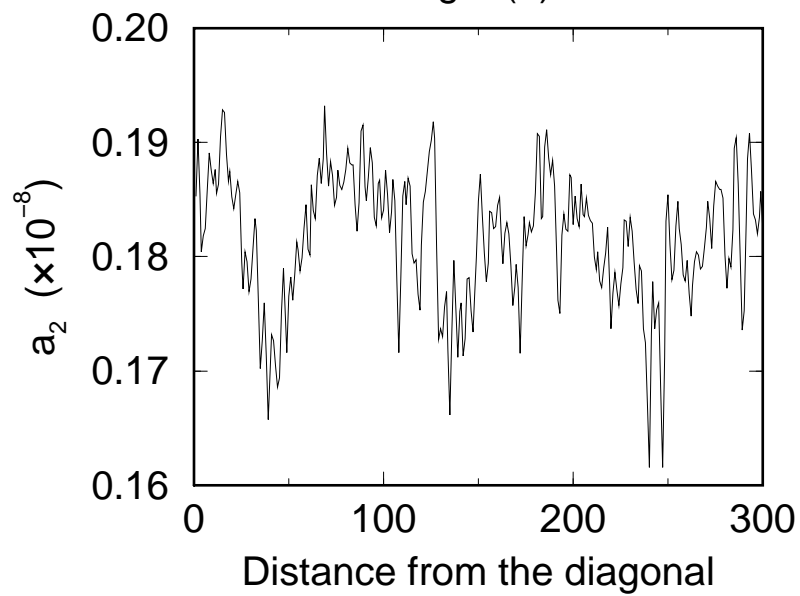
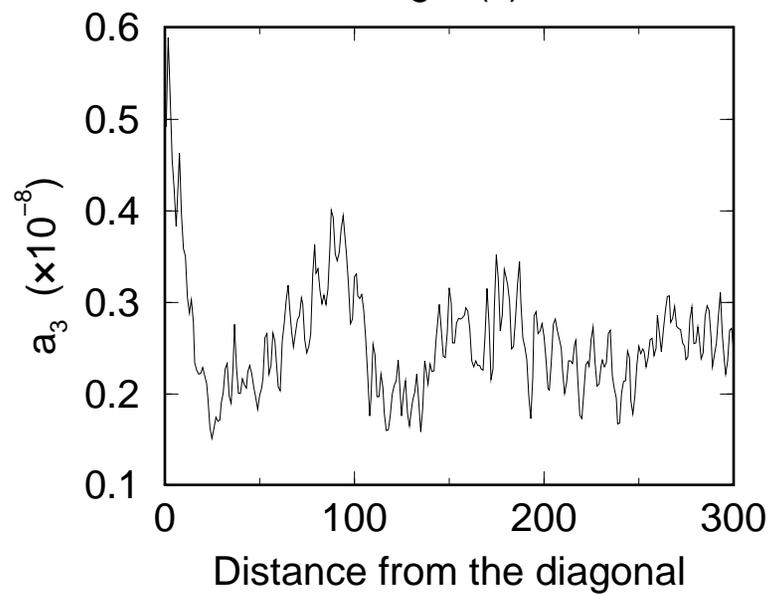
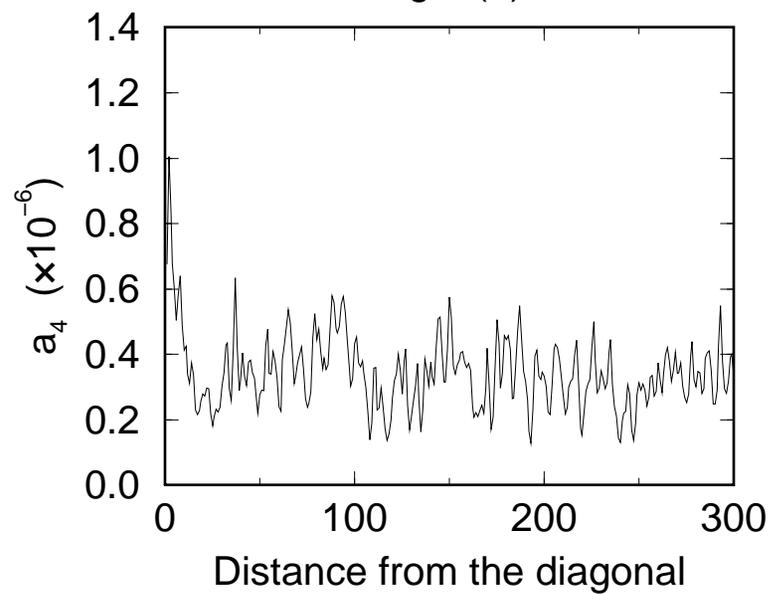
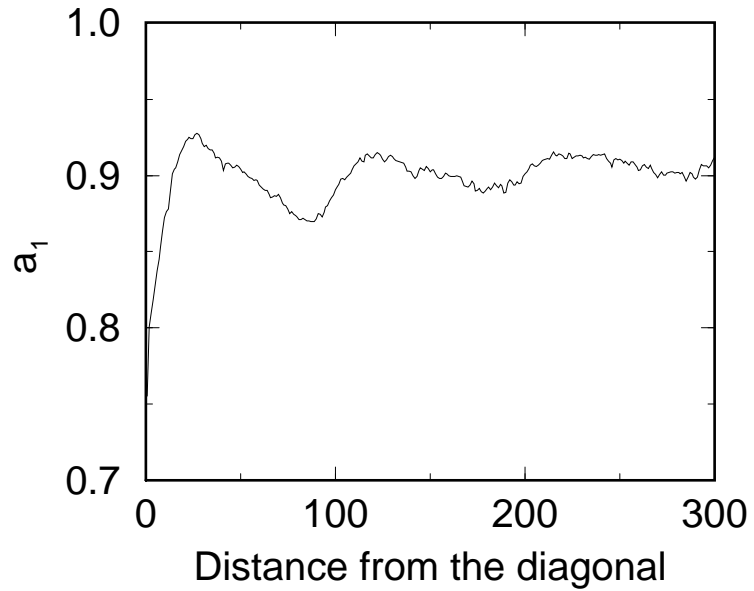
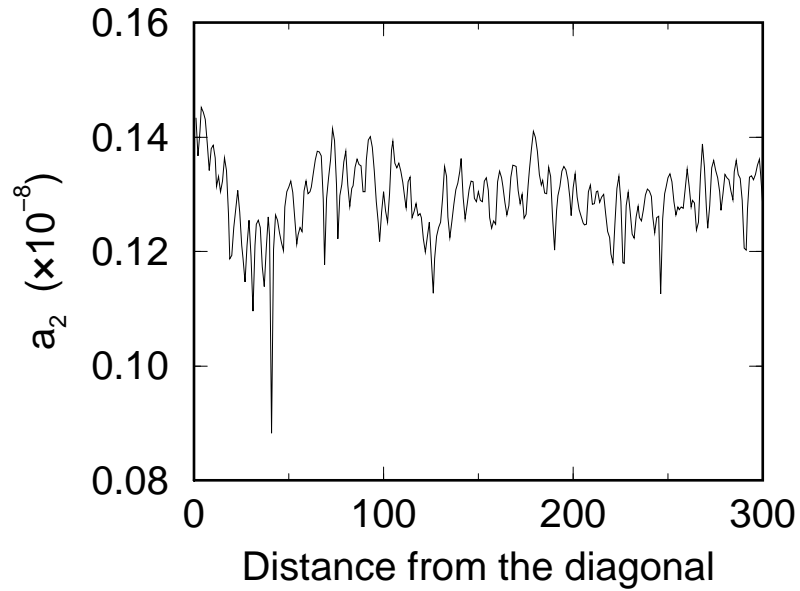
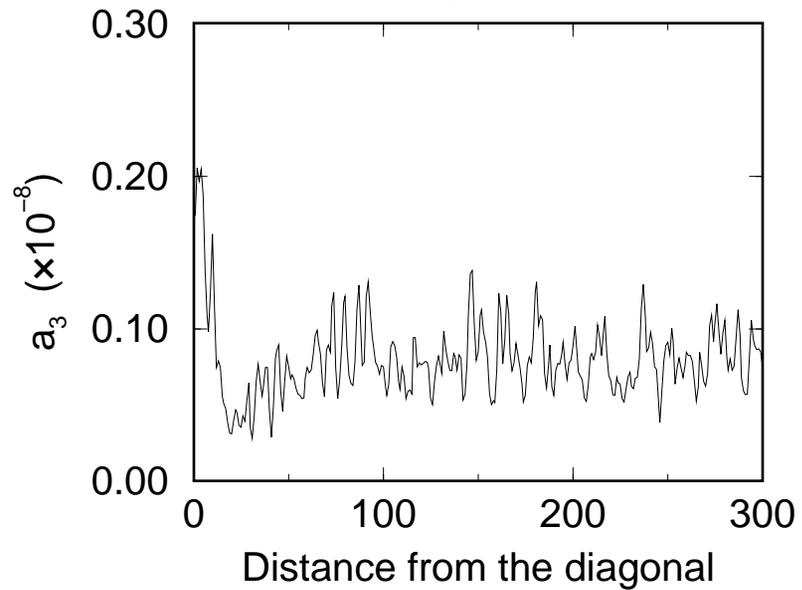
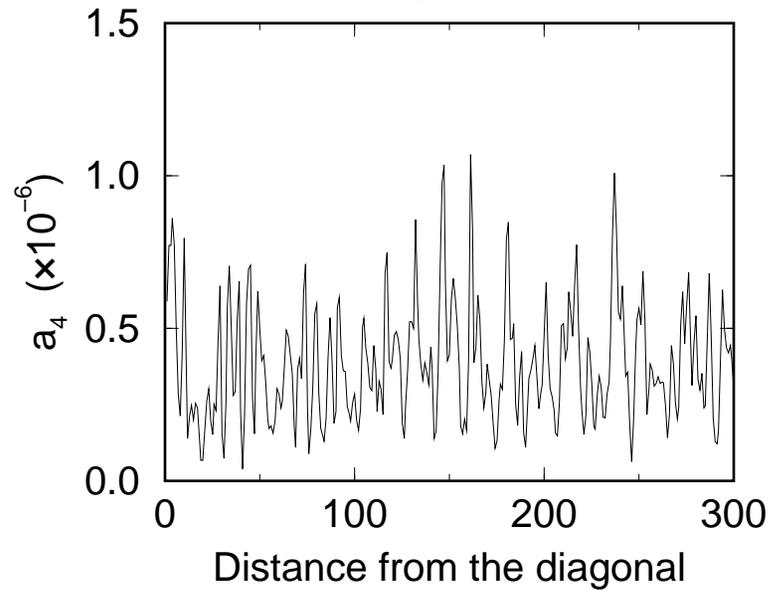
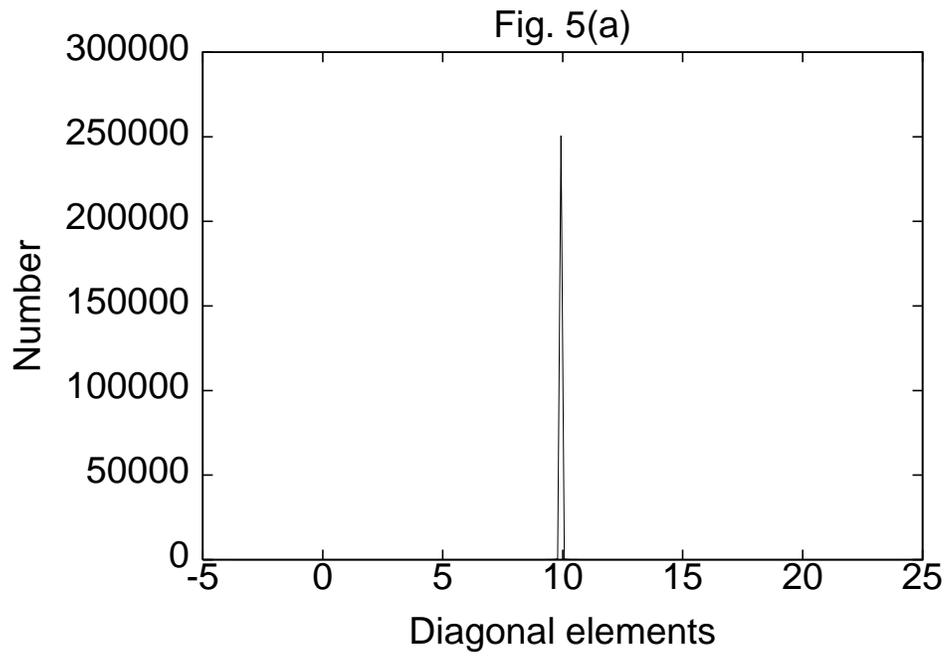
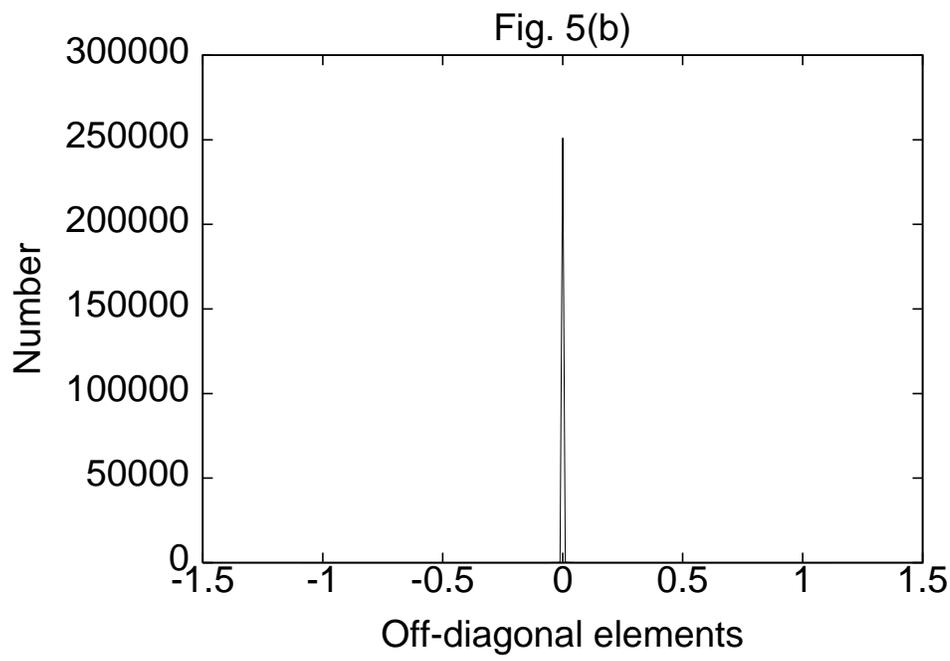
\end{document}